\documentclass{PoS}

\title{Model independent results for nucleon structure}

\ShortTitle{Model independent results for nucleon structure}

\author{\speaker{Gerald A. MILLER}%
         \thanks{This work is partially supported by the USDOE.}\\
        University of Washington, USA\\
        E-mail: \email{miller@phys.washington.edu}}

\abstract{I review recent results regarding the nucleon charge  and magnetization densities as well as the shape of the nucleon.   First,  some phenomenolgical considerations that 
show that the shape of the proton is not round are discussed. Then 
 model independent results regarding the neutron and proton charge density, and the proton magnetization density
are presented. Finally, I show how the spin-dependent densities that reveal  the  shape of the proton can be measured via their relation with transverse momentum distributions. The present  work is made possible by the recent 
tremendous experimental 
progress made at Bates, Mainz and Jefferson Laboratory.
          }

\FullConference{LIGHT CONE 2008 Relativistic Nuclear and Particle Physics\\
		 July 7-11, 2008\\
		 Mulhouse, France}

\begin{document}


\section{Phenomenology}
\def\bea{\begin{eqnarray}}
\def\eea{\end{eqnarray}}
\def\bfp{{\bf p}}\def\bfR{{\bf R}}
\def\bfb{{\bf b}}\def\bfk{{\bf k}}
\def\bfq {{\bf q}}
\def\poinc{Poincar\'{e} }
\newcommand{\eq}[1]{Eq.~(\ref{#1})}

  The  electromagnetic  form
factors are matrix elements of the current operator,  $J^\mu(x)$, between nucleon states
of different momentum:
\bea
\langle p',\lambda'| J^\mu(0)| p,\lambda\rangle =\bar{u}(p',\lambda')\left(\gamma^\mu F_1(Q^2)+i\frac{\sigma^{\mu\alpha}}
{ 2M}q_\alpha F_2(Q^2)
\right) u(p,\lambda),\eea
where $q_\alpha=p'_\alpha-p_\alpha$,  
$Q^2\equiv -q^2>0$, $M$ is the nucleon mass and the 
light-cone helicities are $\lambda,\lambda'$. 
The Sachs form factors are:
$
G_E(Q^2)\equiv F_1(Q^2)-\frac{Q^2}{ 4M^2}F_2(Q^2), G_M(Q^2)\equiv F_1(Q^2)+F_2(Q^2)$.
The early expectation, based on a simple application of helicity conservation at very 
high values of the momentum transfer, 
was that
the $QF_2(Q^2)/2MF_1\sim {m_{\rm quark}\over Q}\rightarrow {G_E(Q^2)\over G_M(Q^2)}={\rm const},$  which is 
also  obtained from non-relativistic considerations. Thus the
expectation was that, at sufficiently large momentum transfer,  $G_E/G_M$ would be flat and
the $QF_2/F_1$ would fall. However, the reverse was true \cite{Jones:1999rz}. 

It is necessary to comment on the meaning of these form factors. In  the Breit frame, with
$\bfp=-\bfp'$,  $G_E$ is the nucleon helicity flip matrix element of $J^0$.  However,
any probability or density interpretation of $G_E$ 
is spoiled by a non-zero  value of $Q^2$, no matter how small.
The initial and final states have different  momentum, and therefore
relativistically have different wave functions. 
Any attempt to analytically correct for the total momentum  by incorporating
 relativistic corrections in a 
$p^2/m_q^2$ type of expansion  is doomed by the presence of the 
very light current quark mass, $m_q$. 
That is, at small values of $Q^2$, one finds
\bea G_E^n\sim Q^2 (\int d^3r \left(r^2 |\psi|^2 +\frac{C} { m_q^2}\right),\eea
where the first term represents the expected  effect depending on the square of the
wave function and the term
 $C$ represents the unknown boost correction. 

So  to analyze form factors using a model, the model must be relativistic. We chose \cite{Frank:1995pv} 
to use a relativistic model using light front coordinates.  
These useful coordinates involve the use of a ``time'' 
$
x^+=(ct +z)/\sqrt{2}=(x^0+x^3)/\sqrt{2}.$
The corresponding evolution operator is the not the Hamiltonian, $p^0$, but instead
$ p^-=(P^0-p^3)/\sqrt{2}.$ The orthogonal spatial coordinate is 
$x^-= (x^0- x^3)/\sqrt{2}.$
If  one quantizes at  $x^+=0$, then $x^-=\sqrt{2}z$, and this why $x^-$ is thought of as the
spatial variable. The canonically conjugate momentum is given by 
$ p^+= (p^0+  p^3)/\sqrt{2}.$ We note that 
$p_\mu x^\mu=p^-x^++p^+x^--\bfp\cdot\bfb.$
The transverse coordinates perpendicular to the 0 and 3 directions are denoted as
$\bfb$ and $\bfp$. Using these variables allows the separation of center of mass and relative
position variables in a manner similar to that of the usual non-relativistic treatment.
The key feature  is that transverse boosts act like the
non-relatistic boosts. 

We used these variables  to formulate a Poincare invariant,  relativistic constituent quark model in 1995 and predict  the qualitative behavior of  form factors
measured five years later.  Please see Figs.~10, 11 of Ref.~\cite{Frank:1995pv}.
More detailed analysis \cite{Miller:2002qb}showed that the flat nature of $QF_2\over F_1$ results from
 orbital angular momentum of the quarks inherent in the lower component of quark Dirac spinors.

So we had a reasonable model, which includes quark orbital angular momentum. 
I was faced with the challenge of relating  orbital angular momentum to a potential non-spherical shape of the proton.
The notion that the proton might not be   a sphere has its
 impetus  in the
  discovery that 
the spins of quarks and anti-quarks account for only about 30\% of the total angular momentum. 
It seems natural to associate non-zero orbital angular momentum with a non-spherical shape,
 but 
the Wigner-Eckart theorem says  that the proton has no quadrupole moment. In response, 
I introduced spin-dependent densities SDD, and later learned these are  common in condensed matter physics.

\newcommand{\boldsigma}{\mbox{\boldmath $\sigma$}}\newcommand{\boldxi}{\mbox{\boldmath $\xi$}}
\newcommand {\boldgamma}{\mbox{\boldmath$\gamma$}}
\newcommand{\boldtau}{\mbox{\boldmath $\tau$}}
\newcommand{\bftau}{\mbox{\boldmath $\tau$}}
\newcommand{\bfx}{{\bf x}}
\newcommand{\bfy}{{\bf y}}
\newcommand{\bfP}{{\bf P}}\def\bfS{{\bf S}}
\def\bfq {{\bf q}}\newcommand{\bfxi}{\mbox{\boldmath $\xi$}}
\def\bfs {{\bf s}}\newcommand{\st}{{\scriptscriptstyle T}}\def\bfs {{\bf s}}
\def\bfn {{\bf n}}
\def\bfqp {{\bf q}_\perp}
\def\bfK{{\bf K}}
\def\bfKp{{\bf K}_\perp}
\def\bfL{{\bf L}}
\def\bfk{{\bf k}}
\def\bfp{{\bf p}}  
\newcommand{\bfkap}{\mbox{\boldmath $\kappa$}} 
\def\bfr{{\bf r}} 
\def\bfy{{\bf y}} 
\def\bfx{{\bf x}} 
\def\be{\begin{equation}}
 \def \ee{\end{equation}}
\def\bea{\begin{eqnarray}}
  \def\eea{\end{eqnarray}}
\newcommand{\eqn} {Eq.~(\ref )}
\newcommand{\bb}{\langle}
\newcommand{\kk}{\rangle}
\newcommand{\bk}[4]{\bb #1\,#2 \!\mid\! #3\,#4 \kk}
\newcommand{\kb}[4]{\mid\!#1\,#2 \!\mid}

\def\notp{{\not\! p}}
\def\notk{{\not\! k}}
\def\up{{\uparrow}}
\def\down{{\downarrow}}
\def\bfb{{\bf b}}

\def\poinc{Poincar\'{e} }
\def\bfq {{\bf q}}
\def\bfK{{\bf K}}
\def\bfL{{\bf L}}
\def\bfk{{\bf k}}
\def\bfp{{\bf p}}  
\def\be{\begin{equation}}
 \def \ee{\end{equation}}
\def\bea{\begin{eqnarray}}
  \def\eea{\end{eqnarray}}
\def\eqn {Eq.~(\ref )}

\newcommand{\kx}[2]{\mid\! #1\,#2 \kk}
\def\notp{{\not\! p}}
\def\notk{{\not\! k}}
\def\up{{\uparrow}}
\def\down{{\downarrow}}
\def\bfb{{\bf b}}

\vskip0.05cm\noindent
{\bf Spin-dependent density operators}
We 
interpret  orbital angular momentum 
 in terms of the shapes of the 
proton exhibited 
through  the rest-frame ground-state 
matrix elements of spin-dependent
density operators \cite{Miller:2003sa}.
The  density operator 
 is 
$ 
\widehat{\rho}(\bfr)= \sum_i 
\delta(\bfr-\bfr_i), $ 
where $\bfr_i$ is the position operator 
of the $i$'th particle. 
For  particles of  spin 1/2 one can measure
the {\it combined}  probability 
that  a particle is at a given position $\bfr$ and has a 
spin in an arbitrary, fixed direction specified by a unit vector
 $\bfn.$ The  coordinate-space spin-dependent density SDD operator
is 
$ \widehat{\rho}(\bfr,\bfn)= \sum_i
\delta(\bfr-\bfr_i){1\over2}(1+\boldsigma_i\cdot\bfn).$

To understand the connection between the  
 spin-dependent density and 
orbital angular momentum, 
  consider a first  example of a
 single charged particle
moving in a fixed, rotationally-invariant potential in an energy eigenstate
$|\Psi_{1,1/2,s}\rangle$ 
of quantum
numbers: $l=1,j=1/2$, polarized in the  direction $\widehat{\bfs}$
 and radial wave function $R(r_p)$. The wave function can be written
as 
$(\bfr_p| \Psi_{1,1/2,s}\rangle=R(r_p)\boldsigma\cdot\hat{\bfr}_p|s\rangle. $ 
The ordinary density. 
$\rho(r)=\langle\Psi_{1,1/2,s}|\delta(\bfr-\bfr_p)|
\Psi_{1,1/2,s}\rangle=R^2(r)$, a spherically symmetric result because  the 
effects of the Pauli spin operator square to unity. 
But 
the matrix element of 
the SDD 
is more interesting:
\bea \rho(\bfr,\bfn) 
={R^2(r)\over 2}\bb  
\widehat{s}\vert\boldsigma\cdot \hat{\bfr}(1+\boldsigma\cdot \hat{\bfn})\boldsigma\cdot\hat{\bfr}
\vert \widehat{s}\kk.\eea
The magnetic  quantum 
defines an axis, $\bfs$  
and
 the direction of vectors can be represented in terms of this axis:
$\hat{\bfs}\cdot\hat{\bfr}=\cos\theta$. Suppose  $\hat{\bfn}$
 is either parallel or anti-parallel to
the direction of the proton angular momentum  vector 
$\hat\bfs$. Then 
$ \rho(\bfr,{\bfn}=\hat{\bfs})={R^2(r)}\cos^2\theta,\;
 \rho(\bfr,{\bfn}=-\hat{\bfs}  )={R^2(r)}\sin^2\theta$, and
 the non-spherical shape is exhibited. 
The average of these
two cases   is  
a spherical shape. 

We  computed momentum-space  SDDs  \cite{Miller:2003sa}
using our model\cite{Frank:1995pv} model and obtained a variety of unusual shapes. See Figs.~2,3
 of  \cite{Miller:2003sa}.
The logic is that using  a model wave function, in rough agreement with data for form factors, 
leads to a non-spherical shape of the SDD. However, there was no direct connection between
experiment and the shapes. 
We shall return to issue  after discussing other model independent results for  
charge and magnetization densities.

\section{Model independent neutron charge density}
The neutron has no net charge, but the charge density 
need not vanish. So we can ask, ``Is the central charge density negative or positive?''.
Long-standing existing answers are based on models For example,. 
the neutron can make a spontaneous quantum transition to a $p\pi^-$ state
\cite{cbm}. The  light pion  spreads out over a larger region of space than
the proton to cause a negative  charge density  at the edge of the neutron and positive one at
the center. A similar  result is obtained with the  different one-gluon exchange
force, acting repulsively between  two negatively charged d-quarks.
But enough  about models! 

The starting point is for a model independent analysis is 
the  
 use of  transversely localized nucleon 
states  \cite{soper1,mbimpact,diehl2}:
\bea
\left|p^+,{\bf R}= {\bf 0},
\lambda\right\rangle
\equiv {\cal N}\int \frac{d^2{\bf p}}{(2\pi)^2} 
\left|p^+,{\bf p}, \lambda \right\rangle.
\label{eq:loc}
\eea
where $\left|p^+,{\bf p}, \lambda \right\rangle$
are light-cone helicity eigenstates and
${\cal N}$ is a normalization factor.
The  range of integration
in \eq{eq:loc} must be restricted to $|\bfp|\ll p^+$ to maintain the interpretation
of a nucleon moving with well-defined longitudinal momentum\cite{mbimpact}. Thus we use
the infinite momentum frame.

Using  \eq{eq:loc} sets 
the  transverse 
center of momentum of 
a state of  total very large 
momentum $p^+$  to zero, so that
transverse distance $\bfb$ relative to $\bfR$.
can be  defined. 
Thus one  defines a useful combination of quark-field operators \cite{mbimpact}:
\bea
\hat{O}_q(x,{\bf b}) \equiv
\int \frac{dx^-}{4\pi}{q}_+^\dagger
\left(-\frac{x^-}{2},{\bf b} \right) 
q_+\left(\frac{x^-}{2},{\bf b}\right) 
e^{ixp^+x^-},
\label{eq:bperp}
\eea 
where the subscript $+$ denotes the use of independent quark field operators.
The  
impact parameter dependent PDF is defined \cite{mbimpact} as:
\bea
q(x,{\bf b}) \equiv 
\left\langle p^+,{\bf R}= {\bf 0},
\lambda\right|
\hat{O}_q(x,{\bf b})
\left|p^+,{\bf R}= {\bf 0},
\lambda\right\rangle. 
\label{eq:def1}
\eea
This is the basic density that can be obtained, giving the probablitiy of finding a quark at a transverse position
that carries a fraction $x$ of the the protons plus component of momentum.
The use of \eq{eq:loc} in \eq{eq:def1} allows one to show \cite{me} that 
$q(x,{\bf b})$ is the two-dimensional Fourier transform of the GPD $H_q$:
\bea q(x,{\bf b})=\int \frac{d^2q}{ (2\pi)^2}e^{i\;\bfq\cdot\bfb}H_q(\xi=0,x,t=-\bfq^2),\label{ft1}
\eea. 

One obtains a relation between $q(x,{\bf b})$ and the form factor \cite{soper1} by integrating $q(x,{\bf b})$
over all values of $x$. This  sets the value of $x^-$ to 0, so that 
a density appears in the matrix element.
If one also multiplies  by  the quark charge $e_q$ (in units of $e$),
sums over quark flavors,  and uses  the sum rule relating the GPD to the form factor,
 the resulting infinite-momentum-frame  IMF parton
charge density in transverse space
is 
\bea
\rho(b)\equiv \sum_q e_q\int dx\;q(x,{\bf b})=\int \frac{d^2q}{ (2\pi)^2} F_1(Q^2=\bfq^2)e^{i\;\bfq\cdot\bfb}.
\label{rhob}\eea


We
exploit \eq{rhob} by  using  using recent parameterizations \cite{Bradford:2006yz,Kelly:2004hm} of measured form factors
to determine $\rho(b)$.
The charge densities of the proton and neutron are shown in Fig.~1 of \cite{me}. The  
surprising feature is the negative central value of the neutron
charge density. This results from the negative definite nature of  $F_1$ \cite{me}. 
The neutron charge density has interesting features, as shown in Fig.~2 of \cite{me} which
displays the quantity $b\rho(b)$. It is the integral of this quantity that integrates to 0.
The neutron charge density is negative at the center, positive in the middle, and again
negative  at the outer edge. The medium-ranged positive charge density is sandwiched by
inner and outer regions of negative charge. This interesting behavior needs to be better understood.

One   gains information about  individual $u$ and $d$   quark densities by invoking charge symmetry 
 \cite{mycsb} so that
 $u,d$ densities in the proton are the same as  $d,u$ densities in the neutron.
The results, \cite{me} 
are that the central up quark density 
is larger than that of the down quark by about 30\%.

\bigskip \noindent{\bf Proton Magnetization Density}
We recently showed  \cite{Miller:2007kt},
that the two-dimensional Fourier transform of the
Pauli form factor $F_2$ plays the role of the maagnetization density.

\section{Measuring the   Non-Spherical Shape of the Nucleon}
While  the 
matrix elements of the spin-density operator 
  reveal highly non-spherical densities, 
 experimentally determining  the proton's non-spherical nature
has remained a challenge. 
Here we   explain  how
matrix elements of the spin-dependent density may be measured using their close connection with       
transverse momentum dependent
parton densities  \cite{Miller:2007ae}. 

The field-theoretic version of the spin-dependent charge 
 density operator
 is a generalization of the operator defined in Ref.~\cite {Miller:2003sa}:
\bea&&\widehat{\rho}_{\rm REL}(\bfK,\bfn)=\int {d^3\xi\over(2\pi)^3} 
e^{-i\bfK\cdot\bfxi}
\left. \bar{\psi}(0)\gamma^0
(1+\boldgamma\cdot\bfn\gamma_5){\cal L}(0,\xi;\;{\rm path})
\psi({\bfxi})\right|_{t=\xi^0=0},
\label{qftrel}\eea 
where $\psi$ is a quark field operator and flavor indices are omitted. 
The quark  field
operators are evaluated at equal time and accompanied by a path-ordered 
exponential link operator
$ {\cal L}((0,\xi;\;{\rm path})$ 
needed for color-gauge invariance.  This matrix element give the probability of finding a quark of three 
momentum
$\bfK$ and spin direction $\bfn$. It depends on the direction of three vectors $\bfK,\bfn$ and the direction of the
spin polarization. Conpared with the  orginally introduced SDD, there is an extra $\gamma^0$ in front of the term that depends on $\bfn$.

Measuring the spin-dependent densities of  requires  that
 the system be  probed with  
 identical  initial and final states. 
But this condition also 
 enters in measurements of  both ordinary and transverse-momentum-dependent
TMD parton distributions. 
The latter  
\cite{Mulders:1995dh} are:
\begin{eqnarray}
\Phi^{[\Gamma]}(x,\bfK) & = &
\left. \int \frac{d\xi^-d^2\xi_\st}{2\,(2\pi)^3} 
\ e^{iK\cdot \xi}
\,\langle P,S \vert \overline \psi (0)\,\Gamma\,{\cal L}(0,\xi;n_-)
\,\psi(\xi) \vert P,S \rangle \right|_{\xi^+ = 0}, \label{projection}
\end{eqnarray}
where the specific 
path $n_-$ is  that of their  Appendix B.
The functions $\Phi^{[\Gamma]}$ depend on the fractional momentum
$x$ = $K^+/P^+$, the trasnverse momentum $\bfK$, and  the proton spin direction.
The operator $\Gamma$ can be any Dirac operator. 
In particular, the shape of the proton is revealed \cite{Miller:2007ae} through
\bea \Phi^{[i\sigma^{i+}\gamma_5]}(x,\bfK)=S_T^ih_1(x,K^2)+{(K^iK^j-{1\over2}\delta_{ij}\over M^2}h_{1T^\perp}(x,K^2).\eea 

It is therefore tempting to 
try to associate an SDD  such as that  of \eq{qftrel}
with TMDs, but one difference  is  essential. 
Parton density operators \eq{projection} 
depend on  quark-field operators defined
at a fixed  light cone time  $\xi^+=\xi^3+\xi^0=0$ while
our SDD 
is  an  equal-time, $\xi^0=0$, correlation 
function.
However, 
a relation
between the two sets of operators is obtained \cite{Miller:2007ae}
by 
  integrating  the TMD over all values of  $x$ setting $\xi^-$ to zero, 
and integrating \eq{qftrel}
over all values of $K_z$ so that $\xi^3=0$. After integration, 
$\xi^\pm=0$ for both functions.
Computed models for the transverse spin dependent densities are shown in Figs.~1,2 of \cite{Miller:2007ae}.

The term  ${h}^\perp_{1T}$ 
causes  distinctive experimental signatures in 
semi-inclusive leptoproduction hadron production experiments  see the list of references  in \cite{Miller:2007ae}.
In each of these cases, the momentum of the virtual photon and its
vector nature provide the analogue of  the  vector $\bfn$  
needed
to define the spin-dependent density. The hadronic  transverse momentum provides
the third, $\bfK_T$.

Of special interest is the reaction $ep\uparrow\rightarrow e' \pi X$
Here the  term ${h}^\perp_{1T}$  causes a distinctive oscillatory 
 dependence
on the angle $3\phi-\phi_{S_{1}},$ \cite{Boer:1997nt} where 
$\phi$ is the angle between  the momentum of the outgoing lepton and the reaction 
plane in the lepton center of mass frame, and $\phi_{S_{1}}$ denotes the direction
of polarization with  respect to the reaction plane.

It is very exciting that experiments  planned  at 
Jefferson Laboratory aim to specifically measure $h_{1T}^\perp$ \cite{harut} and therefore
determine whether or not the  proton is  round.  
We also note that
the non-spherical shape of the nucleon has been established in lattice QCD by computing 
 appropriate moments of  impact parameter dependent gpds. See the talk
of Zanotti in this workshop.

 Our summary of SDDs is that these are closely related to TMDs. If $h_{1T}^\perp$ is not 0, the proton is not
round. Experiment can show that the proton is not round.

\end{document}